\documentstyle[prd,aps,preprint,psfig]{revtex}

\begin{document}
\draft
\title{Isocurvature Fluctuations of the M-theory Axion in a
    Hybrid Inflation Model}
\author{Toshiyuki Kanazawa}
\address{Department of Physics, School of Science, The University of
  Tokyo, Tokyo 113-0033, Japan}
\author{M. Kawasaki}
\address{Institute for Cosmic Ray Research, The University of Tokyo,
  Tanashi 188-0002, Japan}
\author{Naoshi Sugiyama} 
\address{Department of Physics, Kyoto University, Kyoto 606-8502, Japan}
\author{T. Yanagida}
\address{Department of Physics, School of Science, The University of
  Tokyo, Tokyo 113-0033, Japan}
\maketitle

\begin{abstract}
  The M-theory, the strong-coupling heterotic string theory, presents
  various interesting new phenomenologies. The M-theory bulk axion is
  one of these. The decay constant in this context is estimated as
  $F_a\simeq 10^{16}$~GeV. Direct searches for the M-theory axion seem
  impossible because of the large decay constant. However, we point
  out that large isocurvature fluctuations of the M-theory axion are
  obtained in a hybrid inflation model, which will most likely be
  detectable in future satellite experiments on anisotropies of cosmic
  microwave background radiation.
\end{abstract}




\section{Introduction}

It is widely believed that nonperturbative effects play a central role
in describing the real world in superstring theories. Horava and
Witten~\cite{Horava-Witten} have argued that the strong-coupling
heterotic string theory is dual to an M-theory compactified on ${\bf
  S}^1/{\bf Z}_2$. The low-energy limit of this M-theory is well
described by eleven-dimensional supergravity, where the fundamental
energy scale is the eleven-dimensional Planck mass $M_{11}\simeq
6\times 10^{16}$~GeV, rather than the four-dimensional one $M_4\simeq
2\times 10^{18}$~GeV~\cite{Witten}. The four-dimensional Planck mass
$M_4$ is only an effective parameter appearing at low energies.

This M-theory description of strong-coupling heterotic string theory
leads to various interesting new phenomenologies. Banks and
Dine~\cite{Banks-Dine-1} have pointed out that the bulk moduli fields
provide axion candidates in the M-theory compactified further on a
Calabi-Yau manifold, and some of the axions survive at low energies,
since world-sheet instanton effects are suppressed owing to the large
compactification radius in the string tension unit.  One of the string
axions may acquire its mass dominantly from the QCD anomaly, and it
plays the role of the Peccei-Quinn axion~\cite{Peccei-Quinn} in
solving the strong $CP$ problem.

The decay constant $F_a$ of the M-theory axion is estimated
as~\cite{Banks-Dine-1,Choi}
\begin{equation}
  F_a\simeq 10^{16} {\rm GeV}.
\end{equation}
This value greatly violates the constraint $10^{10} {\rm GeV}\lesssim
F_a\lesssim 10^{12}$~GeV, derived in standard
cosmology~\cite{Kolb-Turner}. However, this problem may be solved by
late-time entropy production through decays of moduli
fields~\cite{Kawasaki-Moroi-Yanagida-1,Banks-Dine-2} or a thermal
inflaton field~\cite{Lyth-Stewart}. Since a very low reheating
temperature, such as $T_R\simeq 1$ - $10$MeV, is required to solve the
above problem, it is hard to imagine a consistent production of the
lightest supersymmetric (SUSY) particles, and they cannot be the dark
matter~\cite{Kawasaki-Moroi-Yanagida-2}. Thus, the M-theory axion
seems the most plausible candidate for the dark matter in our
universe.

If the M-theory axion is indeed the dark matter, fluctuations of the
axion density should consist of mixtures of adiabatic and isocurvature
modes in general~\cite{Turner,Lyth,ksy}. In this paper we show that a
hybrid inflation model proposed by Linde and
Riotto~\cite{Linde-Riotto} naturally produces isocurvature
fluctuations of the axion comparable to the adiabatic fluctuations
that are in the accessible range of future satellite experiments on
anisotropies of the cosmic microwave background radiation (CMB). The
present analysis, therefore, confirms the previous
proposal~\cite{Kawasaki-Yanagida}.

\section{Hybrid inflation model}

Let us now discuss the hybrid inflation model~\cite{Linde-Riotto},
which contains two kinds of superfields. One of these fields is $S(x,\ 
\theta)$ and the others are $\psi(x,\ \theta)$ and $\bar{\psi}(x,\ 
\theta)$. The model is based on $U(1)_R$ symmetry under which
$S(\theta)\to e^{2i\alpha}S(\theta e^{-i\alpha})$ and $\psi(\theta)
\bar{\psi}(\theta)\to \psi(\theta e^{-i\alpha}) \bar{\psi}(\theta
e^{-i\alpha})$. The superpotential is then given by
\begin{equation}
    W=S(-\mu^2+\kappa \bar{\psi}\psi),
\end{equation}
where $\mu$ is a mass scale and $\kappa$ a coupling constant. The
scalar potential obtained from this superpotential, in the global SUSY
limit, is
\begin{equation}
  V = \left|-\mu^2+\kappa \bar{\psi} \psi \right|^2+\kappa^2\left| S
  \right|^2 \left(\left|\psi \right|^2+ \left|\bar{\psi}\right|^2
  \right)+D{\rm - terms},
\end{equation}
where scalar components of the superfields are denoted by the same
symbols as the corresponding superfields. The potential minimum,
\begin{equation}
  \langle S \rangle=0,\ \langle \psi \rangle
  \langle\bar{\psi}\rangle=\frac{\mu^2}\kappa,\ \left|
    \langle\psi\rangle \right|=\left| \langle \bar{\psi} \rangle
  \right|,
\end{equation}
lies in the $D$-flat direction $\left|\psi \right| = |
\bar{\psi}|$.\footnote{
  We have assumed $U(1)$ gauge symmetry, where $\psi(x,\ \theta)$ and
  $\bar{\psi}(x,\ \theta)$ have charges opposite of $U(1)$, so that
  the $\psi\bar{\psi}$ term is allowed.}
By the appropriate gauge and $R$-transformations in this $D$-flat
direction, we can bring the complex $S,\ \psi$ and $\bar{\psi}$ fields
on the real axis:
\begin{equation}
  S\equiv\frac{1}{\sqrt{2}}\sigma,\ \psi=\bar{\psi}\equiv\frac12\phi,
\end{equation}
where $\sigma$ and $\phi$ are canonically normalized real scalar
fields. The potential in the $D$-flat directions then becomes
\begin{equation}\label{D-flat-potential}
  V(\sigma,\ \phi) = \left(-\mu^2 + \frac14\kappa\phi^2\right)^2 +
  \frac14 \kappa^2\sigma^2\phi^2,
\end{equation}
and the absolute potential minimum appears at $\sigma=0,\ 
\phi=\bar{\phi}=\mu/\sqrt{\kappa}$. However, for $\sigma > \sigma_c
\equiv \sqrt{2}\mu/\sqrt{\kappa}$, the potential has a minimum at
$\phi=0$. The potential given by Eq.(\ref{D-flat-potential}) for $\phi
= 0$ is exactly flat in the $\sigma$ direction. The one-loop corrected
effective potential (along the inflationary trajectory $\sigma >
\sigma_c$ with $\phi = 0$) is given by~\cite{Dvali}
\begin{eqnarray}\label{one-loop-correction}
  V_{\rm one-loop}&=& \frac{\kappa^2}{128 \pi^2} \left[ (\kappa
    \sigma^2 - 2\mu^2)^2 \ln{\frac{\kappa \sigma^2 -
        2\mu^2}{\Lambda^2}}\right.\nonumber\\& +&\left.  (\kappa
    \sigma^2 + 2\mu^2)^2 \ln\frac{\kappa \sigma^2 + 2\mu^2}{\Lambda^2}
    - 2 \kappa^2\sigma^4 \ln {\frac{\kappa \sigma^2}{\Lambda^2}}
  \right],
\end{eqnarray}
where $\Lambda$ indicates the renormalization scale.

Next, let us consider the supergravity (SUGRA) effects on the scalar
potential (ignoring the one-loop corrections calculated above). The
$R$-invariant K\"ahler potential is given by~\cite{Panagiotakopoulos}
\begin{equation}
  K(S, \psi,\bar{\psi}) = \left| S \right|^2 + \left| \psi \right|^2 +
  \left| \bar{\psi} \right|^2-\frac\beta4 \left| S \right|^4 +\cdots,
\end{equation}
where the ellipsis denotes higher order terms, which we neglect in the
present analysis. Here, we set the gravitational scale $\sim 2.4\times
10^{18}$~GeV equal to unity. Then, the scalar potential
becomes~\cite{Nilles}
\begin{eqnarray}
  V(\sigma,\ \phi)&=&\exp\left( \frac{\sigma^2}2 -
    \frac\beta{16}\sigma^4 + \frac{\phi^2}2 \right)
  \left[\frac14\kappa^2 \phi^2 \sigma^2\left (
      1+\frac{\phi^2}4-\frac{\mu^2}{\kappa}\right)\right.\nonumber\\ 
  &&+ \left.\left( 1+ \frac{\beta-1}2 \sigma^2 + \frac{\beta^2 + \beta
        + 1}4 \sigma^4 \right) \left( -\mu^2+\frac\kappa 4
      \phi^2\right)^2\right].
\end{eqnarray}
As in the global SUSY case, for $\sigma\gtrsim\sigma_c$ the potential
has a minimum at $\phi = 0$. The scalar potential for
$\sigma\gtrsim\sigma_c$ and $\phi=0$ becomes
\begin{equation}\label{sugra-correction}
  V_{\rm SUGRA} = \mu^4\left(1+\frac\beta2\sigma^2+\frac{4\beta^2 +
      7\beta + 2}{16}\sigma^4 + \cdots\right).
\end{equation}

In the first approximation, we assume that the inflaton potential for
$\sigma\gtrsim \sigma_c$ and $\phi=0$ is given by the simple sum of
the one-loop corrections Eq.(\ref{one-loop-correction}) and the SUGRA
potential Eq.~(\ref{sugra-correction}):
\begin{eqnarray}
  V(\sigma)&=&\mu^4\left(1+\frac\beta2\sigma^2+\frac{4\beta^2 + 7\beta
      + 2}{16}\sigma^4 \right)\nonumber\\ 
  &+& \frac{\kappa^2}{128 \pi^2} \left[ (\kappa \sigma^2 - 2\mu^2)^2
    \ln{\frac{\kappa \sigma^2 - 2\mu^2}{\Lambda^2}}\right.\nonumber\\ 
  &+&\left.  (\kappa \sigma^2 + 2\mu^2)^2 \ln\frac{\kappa \sigma^2 +
      2\mu^2}{\Lambda^2} - 2 \kappa^2\sigma^4 \ln {\frac{\kappa
        \sigma^2}{\Lambda^2}} \right].
\end{eqnarray}
Hereafter, we study the dynamics of the hybrid inflation with this
potential.

We suppose that the inflaton $\sigma$ is chaotically distributed in
space at the Planck time and it happens in some region in space that
$\sigma$ is approximately the gravitational scale and $\phi$ is very
small ($\approx 0$). Then, the inflaton $\sigma$ rolls slowly down the
potential, and the region inflates and dominates the universe
eventually. During the inflation, the potential assumes an almost
constant value. Also the Hubble parameter changes only very slightly
and it is given by $H_I = V^{1/2}/\sqrt3\simeq\mu^2/\sqrt3$. When
$\sigma$ reaches the critical value $\sigma_c$, a phase transition
occurs and the inflation ends. In order to solve the flatness and
horizon problem we need an {\it e}-folding number $N$ $\simeq
60$~\cite{Kolb-Turner}. In addition, the adiabatic density
fluctuations during the inflation should account for the observation
by COBE, which leads to~\cite{Bennett}
\begin{equation}\label{inf-n60}
  \left. \frac{V^{3/2}}{V'}\right|_{N=60}\simeq 5.3\times
  10^{-4}.
\end{equation}

The evolutions for $\sigma$ and $N$ are described by 
\begin{eqnarray}
  \dot{\sigma} & = & -\frac{V'}{3H}, \\
  \dot{N} &= & -H,
\end{eqnarray}
which are numerically integrated with $\sigma|_{N=0}=\sigma_c$ and
Eq.~(\ref{inf-n60}) as boundary conditions.  Though the inflaton
potential is parametrized by three parameters ($\kappa,\ \beta$, and
$\mu$), we can reduce the number of the free parameters from three to
two by using the constraint Eq.~(\ref{inf-n60}). We consider $\mu$ as
a function of $\kappa$ and $\beta$, i.e.  $\mu = \mu(\kappa,\ \beta)$.

\begin{figure}
    \centerline{\psfig{figure=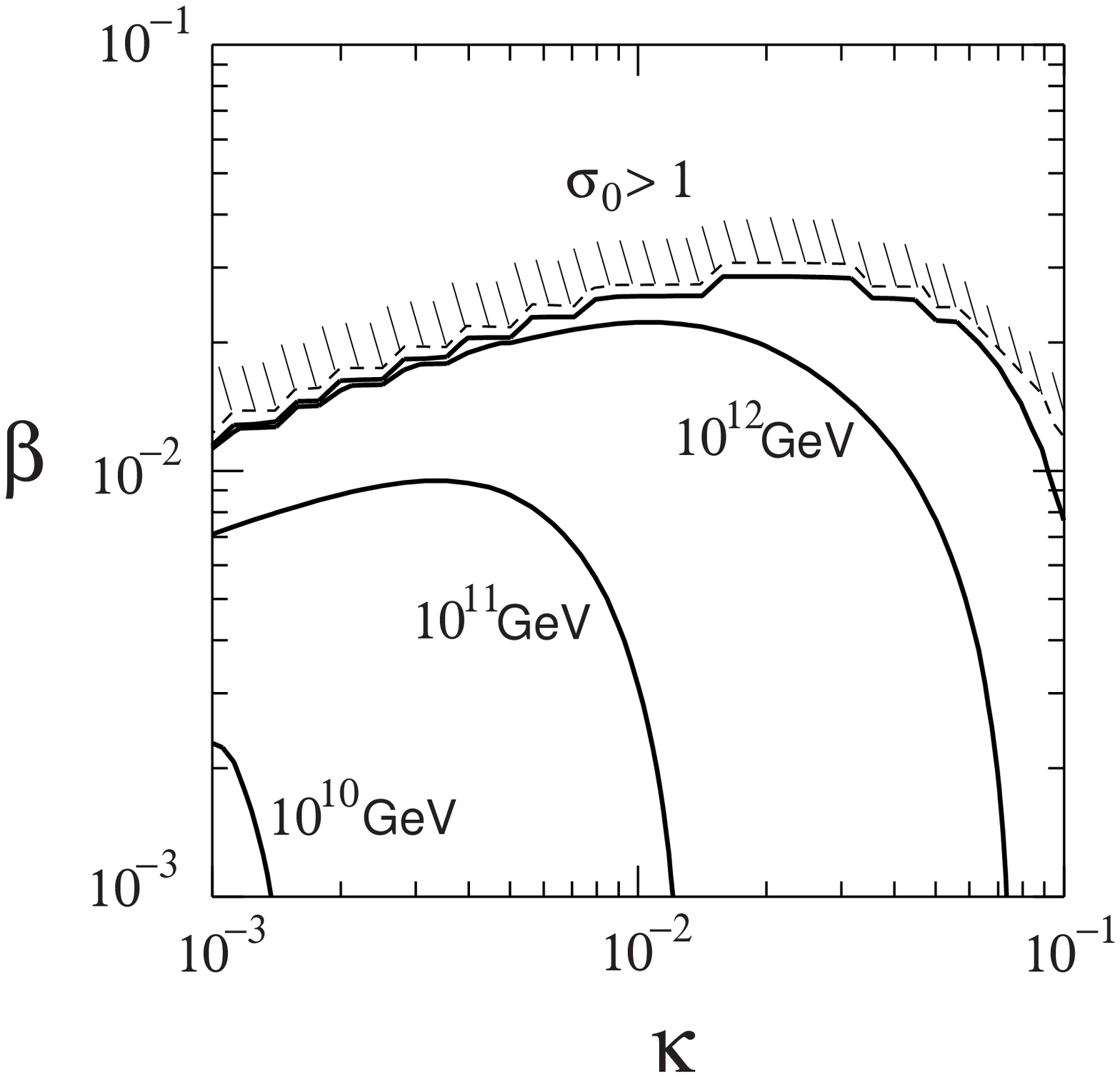,width=14cm}}
    \vspace{1cm}
    \caption{\normalsize
      The Hubble parameter $H_I$ during the inflation, normalized by
      the COBE, ignoring tensor perturbations and isocurvature
      fluctuations. In the region above the dashed line, $\sigma_0$
      exceeds the gravitational scale, $\sigma_0> 1$, and it is
      excluded.}
    \label{fig:no-iso}
\end{figure}

When $\sigma \gtrsim 1$, the slow roll approximation cannot be
maintained. Therefore, if the obtained value of $\sigma|_{N=60}\equiv
\sigma_0$ is larger than the gravitational scale, we should discard
those parameter regions.  Also, one of the attractions of the hybrid
inflation model is that one does not have to invoke extremely small
coupling constants. Thus we assume all of the coupling constants
$\kappa$ and $\beta$ to have values of ${\cal O}(1)$. Here, we choose
$10^{-2}\lesssim \kappa,\ \beta \lesssim 10^{-1}$ as a ``reasonable
parameter region'' (when $\kappa,\ \beta \gtrsim 10^{-1}$, $\sigma_0$
exceeds unity, and $N$ cannot be as large as 60).  The result is shown
in Fig.~\ref{fig:no-iso}, where we plot the Hubble parameter during
the inflation, $H_I=\mu^2/\sqrt3$, as a function of $\kappa$ and
$\beta$.  One may see easily from Fig.~\ref{fig:no-iso} that $H_I$ is
large ($H_I\sim {\cal O}(10^{11-12}{\rm GeV})$) with the reasonable
values of the coupling constants $10^{-2}\lesssim \kappa,\ \beta
\lesssim 10^{-1}$. When $H_I$ has such a large value, the inflation
should generate large isocurvature fluctuations of the axion, if it
exists.  In the above calculations, we have neglected the isocurvature
fluctuations.  Hence, to be consistent, we must take account of the
effects of the isocurvature fluctuations when we normalize the
inflaton potential by COBE; i.e., we must modify Eq.~(\ref{inf-n60}).
We will estimate $H_I$ taking account of the isocurvature effects
later.

\section{Isocurvature fluctuations}

We now evaluate the contribution of the isocurvature fluctuations
assuming the existence of the M-theory axion. Kawasaki, Sugiyama and
Yanagida~\cite{ksy} defined $\alpha$ as the ratio of the initial
entropy perturbation to the adiabatic one when the universe is
radiation dominant, and found it to assume the form~\footnote{
  Our Eq.~(\ref{alpha-old}) is different from Eq.~(5) in
  Ref.~\cite{ksy} by a factor of $1/4$ due to a typographical error
  appearing there.}
\begin{equation}\label{alpha-old}
  \alpha_{\rm KSY}=\frac{9(V')^2}{4H_I^4F_a^2\theta^2} .
\end{equation}
Here, $F_a\theta$ is the initial value of the axion field. In this
paper, we redefine $\alpha$ as the ratio of the {\it present} (the
universe is matter dominant) matter power spectra.  Therefore, in our
notation, $\alpha=1$ implies that the adiabatic and the isocurvature
matter power spectra have the same value in the long wavelength limit.
The old version $\alpha_{\rm KSY}$ in Eq.~(\ref{alpha-old}) is
related to our new version $\alpha$ as\footnote{
  The factor $(2/15)$ comes from the value of the transfer function in
  the long wavelength limit~\cite{Kodama-Sasaki}, and the extra factor
  $(10/9)$ is due to the decay of the gravitational potential at the
  transition from the radiation dominated universe to the matter
  dominated one. }
\begin{equation}\label{alpha-relation}
  \alpha=\left( \frac{2}{15} \right)^2 
  \left( \frac{10}{9} \right)^2 \alpha_{\rm KSY} 
  = \left( \frac{4}{27} \right)^2 \alpha_{\rm KSY} =
  \frac{4(V')^2}{81H_I^4F_a^2\theta^2} .
\end{equation}

The isocurvature fluctuations give a contribution to the CMB
anisotropies which are about six times larger than the adiabatic ones
in the long wavelength limit. At the COBE scales, this factor somehow
decreases, and the precise value depends on $\Omega_0$ and $h$, and is
a monotonically increasing function of these parameters.  Here,
$\Omega_0$ is the ratio of the present energy density to the critical
density, and $h$ is the present Hubble parameter normalized by
$100$~km s$^{-1}$ Mpc$^{-1}$. If we take $\Omega_0\simeq 1$ and
$h\simeq 0.6$, the factor is approximately $\sqrt{30}$.  Therefore,
when we take account of the isocurvature fluctuations, the correct
COBE normalization becomes
\begin{equation}\label{iso-cobe}
  \left. \frac{V^{3/2}}{V'}\right|_{N=60}\simeq \frac{5.3\times
    10^{-4}}{\sqrt{1+30\alpha}},
\end{equation}
rather than Eq.~(\ref{inf-n60}). Here, we have ignored the tensor
perturbations since they are negligibly small in our model. Note that
the spectral index is almost unity in our model.

From Eqs.~(\ref{alpha-old}), (\ref{alpha-relation}), and
  (\ref{iso-cobe}), one may see that 
\begin{equation}\label{fa-alpha}
  \frac{1 + 30\alpha}{\alpha} \simeq 21\times \left(
    \frac{H_I}{10^{12}{\rm GeV}} \right)^{-2} \left( \frac{F_a
      \theta}{10^{16}{\rm GeV}} \right)^{2}.
\end{equation}
For the M-theory axion, the decay constant $F_a$ is estimated as
$F_a\simeq\ 10^{16}$~GeV~\cite{Banks-Dine-1,Choi}. This value is much
larger than the constraint on $F_{a}$, $F_{a} \lesssim 10^{12}$~GeV,
which comes from the requirement that the axion should not overclose
the universe. However, as shown in
Ref.~\cite{Kawasaki-Moroi-Yanagida-1} this constraint is greatly
relaxed if late-time entropy production occurs. In this case, the
unclosure condition for the present universe leads to an upper bound
on $F_a \theta$ :
\begin{equation}
  F_a \theta \lesssim 4.4\times 10^{15}{\rm GeV}.
\end{equation}
Here the reheating temperature after late-time entropy production is
taken as $T_R = 1$~MeV. Thus, a decay constant $F_a\simeq\ 
10^{16}$~GeV of the M-theory axion is allowed if we take $\theta
\lesssim 0.3$, which is not an unnaturally small value. If one takes
$F_{a}\theta \simeq 10^{12}$~GeV as in the standard invisible axion
model, the isocurvature fluctuations become too large. Thus the large
value of $F_{a}$ for the M-theory axion is a rather crucial point in
our mixed fluctuation model.

\begin{figure}
    \centerline{\psfig{figure=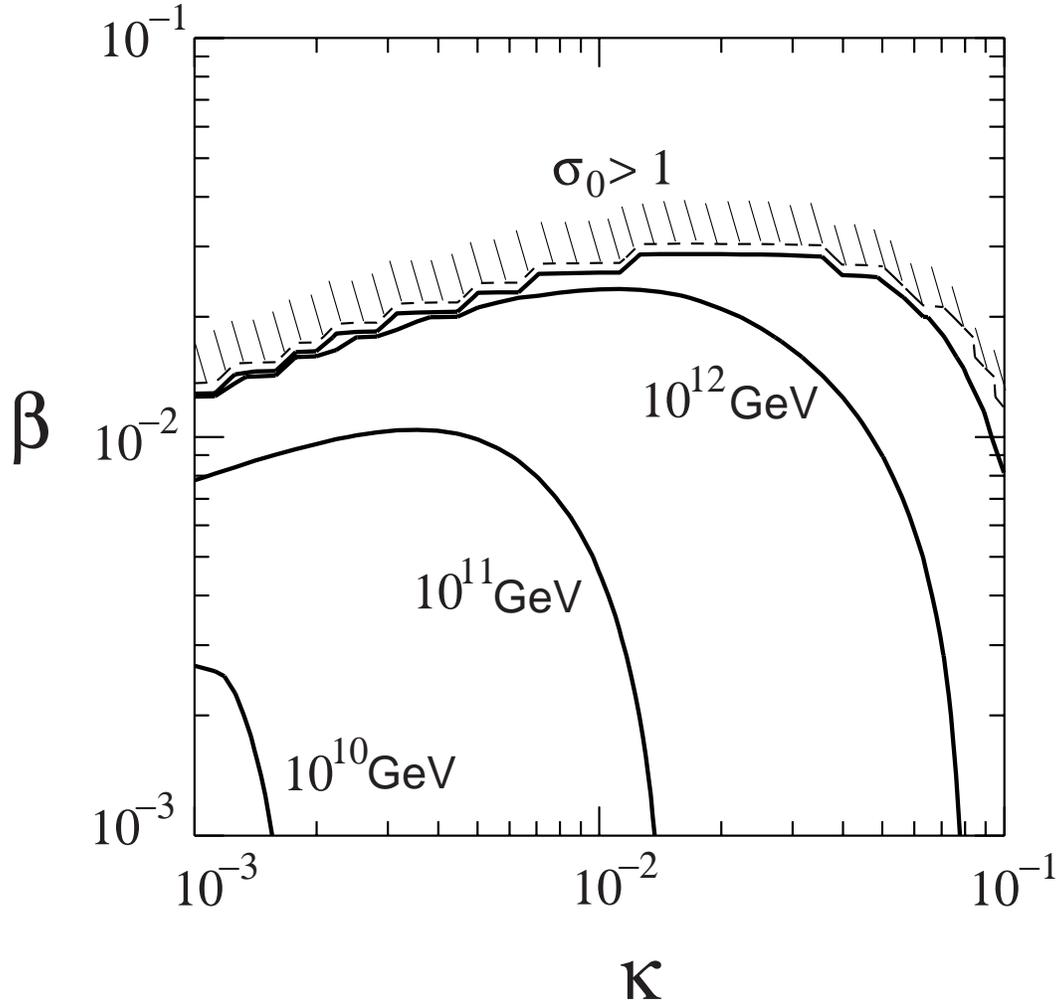,width=14cm}} 
    \vspace{1cm}
    \caption{\normalsize
      The Hubble parameter $H_I$ during the inflation, normalized by
      the COBE for the $\alpha=0.01$ case, ignoring tensor
      perturbations.  In the region above the dashed line, $\sigma_0$
      exceeds the gravitational scale, $\sigma_0> 1$, and it is
      excluded.}
\label{fig:alpha-0.01}
\end{figure}

We are at the point of evaluating $\alpha$, which depends on two
parameters $F_a\theta$ and $H_I$, as seen from Eq.~(\ref{fa-alpha}).
We take $H_I\simeq 10^{11-12}$~GeV, which has been obtained for the
case of $\alpha=0$. We have, however, found that similar values of the
Hubble constant $H_I$ are obtained even for $\alpha\ne 0$, as long as
$\left| \alpha \right|\lesssim 1$. In Fig.~\ref{fig:alpha-0.01} we
plot $H_I$ for $\alpha = 0.01$ as an example. From
Eq.~(\ref{fa-alpha}) we derive $\alpha\gtrsim 0.003$ for $F_a
\theta\lesssim 4\times 10^{15}$~GeV and $H_I\simeq 10^{11-12}$~GeV.

\section{Discussion}

The mixture of isocurvature and adiabatic fluctuations is
astrophysically interesting. Since isocurvature fluctuations yield
anisotropies of the CMB that are six times larger than those caused by
adiabatic fluctuations, mixed fluctuations reduce the amplitude of the
power spectrum if the amplitude is normalized by COBE.  It is well
known that the standard cold dark matter scenario ($\Omega_0=1,
h=0.5$) with COBE-normalized pure adiabatic fluctuations predicts
density fluctuations that are too large on scales of galaxies and
clusters. This problem is avoided if the isocurvature fluctuations are
mixed with adiabatic ones, as is pointed in Ref.~\cite{ksy}.
Furthermore, it can be shown that for the general flat universe
($\lambda_0 +\Omega_0=1$, with $\lambda_0$ the cosmological constant),
the shape and amplitude of the power spectrum are in good agreement
with observations if $\alpha \sim 0.05$~\cite{KKSY}.

\begin{figure}
    \centerline{\psfig{figure=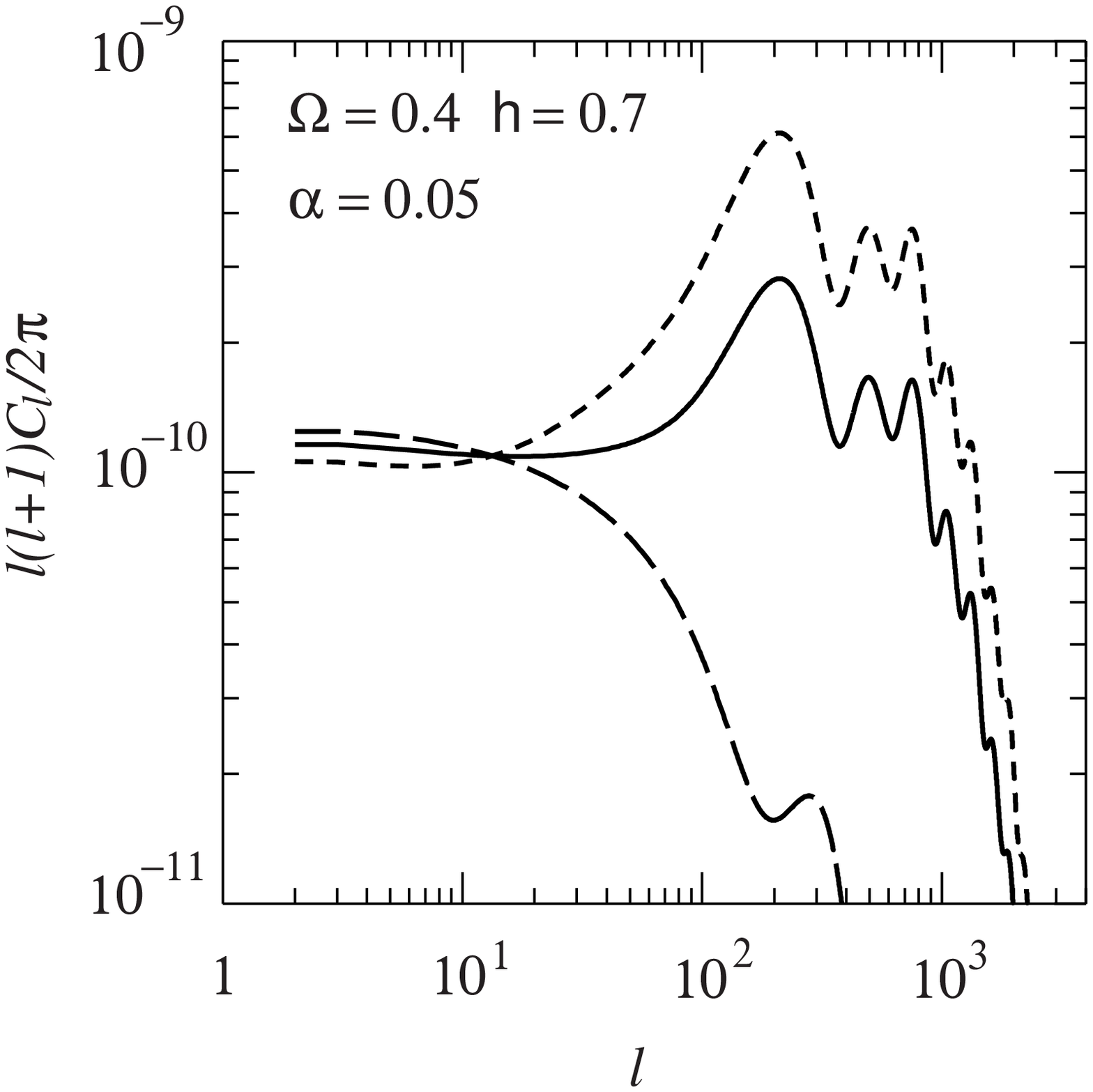,width=14cm}} 
    \vspace{1cm}
    \caption{
      The CMB angular power spectra normalized by the COBE.  Here, we
      have chosen $\Omega_0=0.4,\ h=0.7,\ \lambda_0=0.6$, and
      $\Omega_B h^2=0.015$.  The short dashed line corresponds to the
      pure adiabatic case ($\alpha=0$), the long dashed line to the
      pure isocurvature case ($\alpha=\infty$), and the solid line to
      the case $\alpha = 0.05$.}
\label{fig:mix_sample}
\end{figure}

The CMB anisotropies induced by the isocurvature fluctuations can be
distinguished from those produced by pure adiabatic
fluctuations~\cite{ksy}, because the shapes of the angular power
spectrum of CMB anisotropies are quite different from each other on
small angular scales. The most significant effect of the mixture of
the isocurvature fluctuation is that the acoustic peak in the angular
power spectrum decreases.  In Fig.~\ref{fig:mix_sample} we show the
angular power spectrum for $\alpha=0.05$, $\Omega_0=0.4$ and $h=0.7$
as an example. It is seen that the height of the acoustic peak ($\ell
\sim 200$) in the case of mixed fluctuations is greatly reduced
compared with the pure adiabatic case.

Since the axion decay constant $F_a$ is much higher than
$10^{12}$~GeV, a direct search for the M-theory axion is implausible.
Therefore, observations of CMB anisotropies by future satellite
experiments (MAP~\cite{MAP}, PLANCK~\cite{PLANCK}) are very crucial to
test the M-theory axion hypothesis.

\section*{Acknowledgements}

One of the authors (T.~K.) is grateful to K. Sato for his continuous
encouragement.


\begin{thebibliography}{99}
\bibitem{Horava-Witten} 
    P.~Horava and E.~Witten, Nucl.~Phys.
    {\bf B460}, 506 (1996); Phys.~Rev.~{\bf D54}, 7561 (1996).
\bibitem{Witten} E.~Witten, 
    Nucl.~Phys.~{\bf B471}, 135 (1996).
\bibitem{Banks-Dine-1} T.~Banks and M.~Dine, 
    Nucl.~Phys.~{\bf B479}, 173 (1996); 
    Nucl.~Phys.~{\bf B505}, 445 (1997).
\bibitem{Peccei-Quinn} R.~D.~Peccei and H.~R.~Quinn, 
    Phys. Rev.~Lett.~{\bf 38}, 1440 (1977).
\bibitem{Choi} 
    K.~Choi, Phys.~Rev.~{\bf D56}, 6588 (1997);\\
    K.~Choi and J.~E.~Kim, Phys.~Lett.~{\bf B154}, 393 (1985).
\bibitem{Kolb-Turner} 
    E.~W.~Kolb and M.~S.~Turner, The Early Universe
  (Addison-Wesley, 1990).
\bibitem{Kawasaki-Moroi-Yanagida-1}
    M.~Kawasaki, T.~Moroi and T.~Yanagida, 
    Phys.~Lett.~{\bf B383}, 313 (1996).
\bibitem{Banks-Dine-2} T.~Banks and M.~Dine, 
    Nucl.~Phys.~{\bf B505}, 445 (1997). 
\bibitem{Lyth-Stewart} D.~H.~Lyth and E.~D.~Stewart, 
    Phys.~Rev.~{\bf D53}, 1784 (1996).
\bibitem{Kawasaki-Moroi-Yanagida-2} 
    M.~Kawasaki, T.~Moroi and T.~Yanagida,
    Phys. Lett. {\bf B370}, 52 (1996).
\bibitem{Turner} 
    M.S. Turner and F. Wilczek,
    Phys. Rev. Lett. {\bf 66}, 5 (1991);\\
    A.D. Linde, Phys. Lett. {\bf B259}, 38 (1991).
  \bibitem{Lyth} D.H. Lyth, Phys. Lett. {\bf 236}, 408 (1990);
    Phys.~Rev.~{\bf D45}, 3394 (1992).
\bibitem{ksy} M.~Kawasaki, N.~Sugiyama  and T.~Yanagida,
    Phys.~Rev.~{\bf D54}, 2442 (1996).
\bibitem{Linde-Riotto} A.~Linde and A.~Riotto, 
    Phys.~Rev.~{\bf D56}, 1841 (1997).
\bibitem{Kawasaki-Yanagida} M.~Kawasaki and T.~Yanagida,
    Prog.~Theor.~Phys.~{\bf 97}, 809 (1997).
\bibitem{Dvali} G.~Dvali, Q.~Shafi and R.~Schaefer,
    Phys.~Rev.~Lett.~{\bf 73}, 1886 (1994).
\bibitem{Panagiotakopoulos} C.~Panagiotakopoulos, 
    Phys.~Lett.~{\bf B402}, 257(1997).
\bibitem{Nilles} For a review, H.~P.~Nilles, 
    Phys.~Rep.~{\bf 110}, 1 (1984).
\bibitem{Bennett} C.~L.~Bennett et al., 
    Astrophys.~J.~{\bf 464}, L1 (1996).
\bibitem{Kodama-Sasaki} H.~Kodama and M.~Sasaki,
  Int.~J.~Mod.~Phys.~{\bf A1}, 265 (1986); {\bf A2}, 491 (1987).
\bibitem{KKSY}T. Kanazawa, M. Kawasaki, N.~Sugiyama and T.~Yanagida,
    in preparation. 
\bibitem{MAP}
    http://map.gsfc.nasa.gov/.
\bibitem{PLANCK}
    http://astro.estec.esa.nl/SA-general/Projects/Planck/.
\end{thebibliography}
\end{document}